\documentclass{ws-ijgmmp}

\usepackage{xcolor}
\usepackage[verbose,hypertexnames=false]{hyperref}
\usepackage{soul}
\usepackage{xcolor}
\sethlcolor{yellow}
\hypersetup{colorlinks=false,allbordercolors=blue,pdfborderstyle={/S/U/W 1}}

\begin{document}

\markboth{M.Shokrollahi, B. Mirza, M. Tavakoli Kachi}
{The Shadow of Rotating FJNW Metric}

%
\catchline{}{}{}{}{}
%

\title{The Shadow of Rotating FJNW Metric}

\author{Mina Shokrollahi\textsuperscript{*}}


\address{Department of Physics, Isfahan University of Technology, Isfahan, 84156-83111, Iran.\\
\,
\email{mina.shokrollahi1998@gmail.com}}

\author{Behrouz Mirza\textsuperscript{*}}

\address{Department of Physics, Isfahan University of Technology, Isfahan, 84156-83111, Iran.\\
 b.mirza@iut.ac.ir }

\author{Mahnaz Tavakoli Kachi}

\address{Department of Physics, Isfahan University of Technology, Isfahan, 84156-83111, Iran.\\
	m.tavakoli1399phy@gmail.com }

\footnotetext{*Corresponding authors.} 
\maketitle


\begin{abstract}
	We investigate the astrophysical aspects of the rotating Fisher–Janis–Newman–Winicour (FJNW) metric recently derived in \cite{mirza-stationary}. The Hamilton–Jacobi equation is used to obtain the shadow of the rotating FJNW metric. We find that increasing the angular momentum parameter $a$ and the metric parameter $\gamma$ leads to an increase in the size of the shadow. Finally, we study the deflection angle of light in the rotating FJNW spacetime. 
\end{abstract}




\section{Introduction}

Key milestones validating general relativity (GR) include the detection of gravitational waves and black hole imaging, marking significant advances in astrophysics. In 2015,  LIGO-Virgo collaboration  achieved a historic breakthrough by directly detecting gravitational waves emitted from a black hole collision \cite{ligo-1}.  This discovery opened a new frontier in observational astronomy. Also, in 2019, the Event Horizon Telescope (EHT) collaboration  released the first image of the event horizon of the black hole in galaxy M87. The image revealed the black hole's shadow against its glowing accretion disk, providing strong support for the predictions of GR and offering a glimpse into extreme gravitational conditions \cite{m87-1,m87-2,m87-3,m87-4,m87-5,m87-6,m87-7,m87-8,m87-9,m87-10,m87-11, m87-12,m87-13,m87-14,shadow-additional}. 
\\
Then, in 2022,  the Event Horizon Telescope (EHT) released a second image of the black hole at the center of the Milky Way, Sagittarius A*, providing valuable comparative data and further enhancing our understanding of supermassive black holes in diverse astrophysical environments \cite{sga-1,sga-2,sga-3,sga-4,sga-5}.

Einstein’s theory of general relativity has profoundly influenced our understanding of gravity and spacetime, not only in the context of black holes \cite{bh-1,bh-2,bh-3,bh-4} but also for other exotic objects such as wormholes \cite{wormhole1,wormhole2,wormhole3,wormhole4,wormhole5,wormhole6,wormhole7}. 

The concept of the "black hole shadow" was first articulated by  "James Bardeen" in 1973 \cite{barden}. It describes how a black hole's intense gravitational field bends light, creating a dark region against the bright background of surrounding matter. This shadow corresponds to the silhouette of the event horizon, marking the boundary beyond which nothing can escape. By analyzing the black hole shadow, one can infer key properties such as the mass and spin of the black hole, as well as the geometry of the surrounding spacetime. Moreover, studies of black hole shadows open new avenues for testing gravitational theories and exploring strong-field gravity \cite{grav1,grav3,grav4,grav5,grav6,grav7,grav8,grav9,Ferrari,celestial-0,celestial-1,celestial-2}.

Coordinate-independent approaches, based on Legendre polynomial expansions, Fourier series, and neural networks, have been developed to compare shadow features across a broader range of spacetime geometries \cite{coor-ind-1,coor-ind-2}.    

One of the significant gravitational effects associated with black holes is the deflection of light. Due to the powerful gravitational field of black holes, light beams passing near them deviate from their original trajectories. In 1848, Carl Friedrich Gauss and Pierre Bonnet introduced a geometric method for calculating light deflection angle using the Gauss–Bonnet theorem \cite{sGB}. This theorem provides a rigorous mathematical framework for determining the deflection angle of light beams in curved spacetime. Subsequently, in 2008, Gibbons and Werner developed an alternative and widely used approach for computing the light deflection angle based on this formalism \cite{Gibbons}.
\\
They assumed that both the light source and the observer are located in an asymptotically Minkowskian region. Using the Gauss–Bonnet theorem, they described the behavior of light beams in curved spacetime. However, their method was restricted to spherically symmetric spacetimes and could not be straightforwardly applied to axisymmetric geometries due to the presence of off-diagonal time–space components in the metric. These off-diagonal terms complicate the construction of the optical metric in axisymmetric cases, thereby necessitating a more general and comprehensive approach \cite{new-guassbonnet}.

The detection of black hole images and gravitational-wave signals represent transformative milestones in our understanding of the universe. These achievements validate Einstein’s theory of general relativity and propel the exploration of cosmic phenomena, thereby expanding the horizons of our scientific knowledge \cite{sourceof-GB,def3}.

According to the theory of general relativity, the collapse of massive objects inevitably leads to the formation of a spacetime singularity \cite{general1,general2,general3,general4}. However, such a collapse does not always result in an event horizon. If a singularity forms without an event horizon, it is referred to as a naked singularity. Like a black hole, this region is characterized by extremely high density, where physical quantities diverge, but unlike a black hole, the singularity is not hidden by a horizon. Naked singularities are particularly intriguing because they challenge our understanding of the universe by exposing regions of spacetime that are otherwise concealed. Studying these horizonless objects pushes the boundaries of our knowledge and may provide insights into the fundamental laws of physics, the nature of gravity, and the origins of cosmic phenomena \cite{naked1,naked2,naked3,naked4}.

One of the well-known solutions of Einstein's equations is 
FJNW metric \cite{fisher,j-n-w,Wyman,K.S}.
FJNW metric describes a static naked singularity in a massless scaler field. If the scalar field is set to zero, the metric reduces to the Schwarzschild solution. Therefore, the singularity of  FJNW metric arises due to the existence of a scalar field.  Gravitational lensing of FJNW metrics has been extensively studied in \cite{scalerfield-1,scalerfield-2,gra-lens-bh-naked-2,gra-lens-bh-naked-3,gra-lens-bh-naked-4.5,gra-lens-bh-naked-4.6,shadow-rotating-fjnw, gra-lens-bh-naked-4.7,gra-lens-bh-naked-4.8,gra-lens-bh-naked-5,gra-lens-bh-naked-6, gra-lens-bh-naked-7,gra-lens-bh-naked-8,gra-lens-bh-naked-9,gra-lens-bh-naked-10}. The existence of a shadow, however, is not restricted to spacetimes possessing an event horizon. Naked singularities and other horizonless compact objects may also exhibit well-defined shadow structures, determined by the properties of their null geodesics and photon regions \cite{naked-shadow-1}.

Recently, a three-parameter solution of Einstein's equations in a massless scaler field was introduced in \cite{mirza-static}. The correct form of the rotating FJNW metric for entire parameters' values and a real scalar field was obtained for the first time in \cite{mirza-stationary}. For a review of the earlier wrong solutions that was proposed for the rotating FJNW metric see \cite{wrong-solutions-1}. Additionally, Bogush and Gal'tsov proposed a rotating FJNW solution corresponding to a limited range of parameters and a phantom scalar field in \cite{wronganswer-additional}.

In this paper, we investigate the astrophysical aspects of rotating FJNW metrics, focusing on their shadows and gravitational lensing. The shadow arises from photons bending into unstable orbits, known as light rings, around the dense mass distribution of a black hole or naked singularity. We employ the Hamilton–Jacobi equation to compute the shadow by perturbing the metric. Using this perturbative approach, we analyze the shape of the shadow and its variation for different values of the angular momentum $a$ and the metric parameter $\gamma$.
\\

This paper is organized as follows. In Section \ref{Section2}, we represent the rotating FJNW metric. In Section \ref{Section3}, we derive the geodesic equations for this metric. In Section \ref{Section4}, we investigate the effective potential, while in Section \ref{Section5}, we calculate the shadow of the rotating FJNW metric. In Section \ref{gauss-bonnet-sec}, we review the Gauss–Bonnet theorem and derive the light deflection angle of the rotating FJNW metric. Throughout the paper, we perform the calculations in units where $ c=8 \pi G=1 $.

\section{Rotating FJNW Metrics}\label{Section2}

The Einstein-Hilbert action in the presence of a scalar field can be written as follows
\begin{equation}\label{action-static}
	S= \int d^4 x \, \sqrt{-g} \, (R -\, \partial_{\rho}  \varphi(r) \, \partial^{\rho} \varphi(r)),
\end{equation}
where $R$ is the Ricci scalar, $g$ is the determinant of metric, and $\varphi(r)$ is the scalar field which is assumed to depend only on the radial parameter $r$.
Having the action, the equations of motion can be obtained by varying Eq. (\ref{action-static}) regarding the metric tensor $g_{\mu \nu}$ and the massless scalar field  $\varphi(r)$ as follows
\begin{equation}
	\begin{split}
		&R_{\alpha \beta}=\partial_{\alpha}  \varphi(r) \ \partial_{\beta} \varphi(r), \\ 
		&\triangledown_{\alpha} \, \triangledown^{\alpha} \, \varphi(r)=0,
	\end{split}
\end{equation}
Using the equations above, the static FJNW metric in a massless scaler field, can be written in the following form
\begin{equation}
	ds^2= -f^\gamma \, dt^2+ \frac{dr^2}{f^{\gamma}}+f^{1-\gamma} \, r^2 \, d\theta^2 
	+f^{1-\gamma} \, r^2 \, \sin^2\theta \, d\phi^2,
\end{equation}
where $\gamma$  is an independent parameter, and the function $f$ is defined as
\begin{equation}\label{static-metrics}
	f(r)=1-\frac{2m}{r}.
\end{equation}
Also, the scalar field is given by
\begin{equation}
	\varphi(r)=\sqrt{\frac{1-\gamma^2}{2}}  \, \ln  (1-\frac{2m}{r}).
\end{equation}
For a real scaler field and a positive mass, the condition
$ 0\leq \gamma \leq1$ should be satisfied.
\\
Although several rotating extensions of the FJNW metric have been proposed, none of them satisfy the Einstein field equations \cite{wrong-solutions-1,wronganswer-additional}. The correct form of the rotating solutions was derived in \cite{mirza-stationary}.
\\
The  rotating solution of FJNW metric in the presence of a massless scaler field \cite{mirza-stationary} is given by 
\begin{equation}\label{totalmetricofjnw}
	\begin{split}
		ds^2=&-f \ (dt-\omega \, d\phi)^2+\frac{\sigma^2}{f} \ [e^{2\eta}(x^2-y^2) \\
		&\times (\frac{dx^2}{x^2-1}+\frac{dy^2}{1-y^2})+(x^2-1) \ (1-y^2) \ d\phi^2],
	\end{split}
\end{equation}
where $\sigma$ is a positive constant and the functions $f$, $\omega$, and $e^{2\eta}$ take the following forms, respectively
\begin{equation}\label{para}
	\begin{split}
		&f=\frac{A}{B},\\
		&\omega=-2 \ (a+\sigma \ (\frac{C}{A})),\\
		&e^{2\eta}=\frac{1}{4} \ (1+\frac{m}{\sigma})^2 \  \frac{A}{(x^2-1)^{1+q}} \ (\frac{x^2-1}{x^2-y^2}),
	\end{split}
\end{equation}
where $a$ is a spin parameter,  $m$ is  a mass parameter, and $q=\gamma-1$ is the metric parameter; also,  $A, B$ and $C$ in Eq. (\ref{para}) are defined as follows
\begin{equation}
	\begin{split}
		&A=a_{+} \ a_{-}+b_{+} \ b_{-},\\
		&B=a_{+}^2+b_{+}^2,\\
		&C=(x+1)^{q} \ [x \ (1-y^2) \ (\rho+\xi) \ a+y \ (x^2-1) \ (1-\rho\, \xi) \ b_{+}],
	\end{split}
\end{equation}
where,
\begin{equation}
	\begin{split}
		&a_{\pm}=(x \pm 1)^q \ [x \ (1-\rho\, \xi) \pm (1+\rho\, \xi)],\\
		&b_{\pm}=(x \pm 1)^q \ [y \ (\rho+\xi) \mp (\rho-\xi)],\\
		&\rho=\alpha \ (x^2-1)^{-q} \ (x+y)^{2 \, q},\\
		&\xi=\alpha \ (x^2-1)^{-q} \ (x-y)^{2 \, q},\\
		&\alpha=-\frac{a}{m+\sigma},\\
		&\sigma=\sqrt{m^2-a^2},
	\end{split}
\end{equation}
and the prolate coordinates x and y are defined by
\begin{equation}
	\begin{split}
		&x=\frac{r-m}{\sqrt{m^2-a^2}},\\
		&y=\cos \theta.
	\end{split}
\end{equation}
Also, the scalar field is given  by 
\begin{equation}
	\varphi(r)=\sqrt{\frac{1-\gamma^2}{2}}  \, \ln  (\frac{r-m-\sqrt{m^2-a^2}}{r-m+\sqrt{m^2-a^2}}).
\end{equation}
By setting $\gamma = 1$ ($q=0$), the scalar field becomes zero, and the metric in  Eq. (\ref{totalmetricofjnw}) reduces to the  Kerr metric.
\\
Now, we expand the metric's elements up to the second order of the rotation parameter $a$ and the first order of the parameter $q$ and neglect terms proportional to $ a\, q $. Therefore, in the Boyer–Lindquist coordinates $(t,r,\theta,\phi)$, we write the FJNW metric as follows
\begin{equation}\label{jnw-metric}
	ds^{2}= g_{tt} \, dt^2+g_{rr} \, dr^2+g_{\theta\theta} \, d\theta^2+g_{\phi\phi} \, d\phi^2+2g_{t\phi} \, dt \, d\phi,
\end{equation}
where, the metric coefficients are given by
\begin{equation}
	\begin{split}
		\nonumber
		g_{tt}=&
		-(1 - \frac{2 \, m}{r}) - \frac{
			2 \, m \, a^2 \cos^2\theta }{r^3}-\\
		& q \, (1-\frac{2m}{r}) \,  \ln[1-\frac{2m}{r}],
	\end{split}
\end{equation}\label{coefficient}
\begin{equation}
	\begin{split}
		\nonumber
		g_{rr}=&(1-\frac{2m}{r})^{-1}+\frac{(-1 + (1-\frac{2m}{r}) \cos^2\theta) \,  a^2)}{
			r^2 \, (1-\frac{2m}{r})^2}-\\
		&q \, \frac{ \ln[1 - \frac{2 \, m}{r}]}{(1-\frac{2m}{r})},
	\end{split}
\end{equation}
\begin{equation}
	\begin{split}
		\nonumber
		g_{\theta \theta}=
		r^2 +a^2 \cos^2\theta-
		q \, r^2 \,  \ln[1 -\frac{2 \, m}{r} ],
	\end{split}
\end{equation}
\begin{equation}
	\begin{split}
		\nonumber
		g_{\phi \phi}=&
		r^2 \,  \sin^2 \theta +\frac{(m \, (1-\cos2 \theta)+r) \, a^2 \,  sin^2\theta )}{r}-\\
		&q \, r^2 \,  \sin^2\theta \,  \ln[1-\frac{2m}{r}],
	\end{split}
\end{equation}
\begin{equation}\label{jnw-metric-coefficients}
	\begin{split}
		g_{t\phi}=\frac{2 \, a \,m \ \sin^2\theta}{r}.
	\end{split}
\end{equation}

It should be noted that each rotating FJNW metric component consists of two parts: one that depends on the metric parameter 
$q$ and another that is independent of it. Therefore, by setting $q=0$, the metric reduces to the rotating Kerr solution.

In the present work, the perturbative expansion
provides the required separability of the Hamilton–Jacobi equation (Section \ref{Section3}) and enables the analytical construction of the photon trajectories that
determine the apparent shadow boundary (Section \ref{Section4}). This is consistent with a strong field phenomenon while retaining the leading-order contributions of the rotation parameter $a$ and the scalar-field parameter $q$.

\section{ Equations of Motion}\label{Section3}

In this section, we aim to investigate the motion of massless particles in the rotating FJNW spacetime. Photon trajectories near the chosen compact object are governed by the corresponding equations of motion \cite{Ferrari}.
Since the rotating FJNW spacetime is stationary and axisymmetric, the geodesic motion is characterized by two conserved quantities
\begin{equation}\label{E-L}
	E=-g_{t\mu} \,  \dot{x}^{\mu}=-p_{t},  \quad L=g_{\phi \mu} \, \dot{x}^{\mu}=p_{\phi},
\end{equation}
where $p_{\mu}$ is the conjugate momenta, and $E$
and
$L$
are constants of motion representing the particle’s energy and angular momentum, respectively. Equation (\ref{E-L}) can be readily solved for $\dot{t}$ and $\dot{\phi}$, yielding
\begin{equation}\label{tdot-phidot}
	\begin{split}
		&\dot{t}=\frac{1}{-g_{tt} \, g_{\phi \phi}+g_{t \phi}^2} \, (g_{\phi \phi} \, E+g_{t \phi} \, L),\\
		&\dot{\phi}=\frac{1}{-g_{tt} \, g_{\phi \phi}+g_{t \phi}^2} (-g_{tt} \, L-g_{t \phi} \, E),
	\end{split}
\end{equation}
where the metric coefficients are defined in Eq. (\ref{jnw-metric-coefficients}). 
To derive the remaining components of the equations of motion, we use Hamilton-Jacobi equation as follows 
\begin{equation}\label{Hamilton-Jacobi}
	H(x^{\mu}, \frac{\partial S}{\partial x^{\mu}})+ \frac{\partial S}{\partial \lambda}=0,
\end{equation}
assuming the solution $S(x^{\mu}, \lambda)$, called action, is a complete integral, then 
\begin{equation}\label{s-pmu}
	\frac{\partial S}{\partial x^{\mu}}=p_{\mu},
\end{equation}
Then, using the following three constants of motion $ H=\frac{1}{2} \, g^{\mu\nu} \, p_{\mu} p_{\nu}$, $E$, and $L$, Eq. (\ref{s-pmu})  can be written for photons as follows
\begin{equation}\label{Scobi}
	\frac{\partial S}{\partial \lambda}=-\frac{1}{2} \, g^{\mu \nu} \, \frac{\partial S}{\partial x^{\mu}}\frac{\partial S}{\partial x^{\nu}},
\end{equation}
where $\lambda$ denotes the affine parameter and $g^{\mu \nu}$ is the inverse form of the metric coefficients in Eq. (\ref{jnw-metric-coefficients}) as follows
\begin{equation}\label{inverse1}
	g^{\mu\nu}=\begin{pmatrix}
		g^{tt} & 0 & 0 & g^{t\phi}\\
		0 & {g_{rr}}^ {-1} & 0 & 0 \\
		0 & 0 & {g_{\theta\theta}}^ {-1} & 0\\
		g^{t\phi} & 0 & 0 & g^{\phi\phi}
	\end{pmatrix}
	,
\end{equation}
where
\begin{equation}\label{inverse2}
	\begin{pmatrix}
		g^{tt} & g^{t\phi}\\
		g^{t\phi} & g^{\phi\phi}
	\end{pmatrix}
	=(g_{tt}g_{\phi\phi}-{g_{t\phi}}^{2})  \begin{pmatrix}
		g_{\phi\phi} & -g_{t\phi}\\
		-g_{t\phi} & g_{tt}
	\end{pmatrix}
	,
\end{equation}
Now, using the separation method, Eq. (\ref{Scobi}) for photons can be written as 
\begin{equation}\label{Splus}
	\frac{\partial S}{\partial \lambda}=E^2 \,  g^{tt}+(\frac{dS_{r}}{dr})^{2}\,g^{rr} + (\frac{dS_{\theta}}{d\theta})^{2}\, g^{\theta\theta}+L^2 \,  g^{\phi\phi}- 2\, E \, L \, g^{t\phi},
\end{equation}
where the constants $E$ and $L$ are defined in Eq. (\ref{E-L}). Also,  $ (\frac{dS_{r}}{dr})^{2}$ and $(\frac{dS_{\theta}}{d\theta})^{2}$ are the functions of $r$ and $\theta$, respectively.

To derive the equations governing the radial and angular motion, $\dot{r}$ and $\dot{\theta}$, it is necessary to obtain the radial and angular functions $R(r)$ and $\Theta(\theta)$ from the Hamilton--Jacobi equation given in Eq.~(\ref{Hamilton-Jacobi}). For the exact rotating FJNW spacetime Eq. (\ref{totalmetricofjnw}), the Hamilton--Jacobi equation is not separable into purely radial and angular parts. Therefore, we employ the perturbative form of the rotating FJNW metric introduced in Eq. (\ref{jnw-metric-coefficients}). The metric components are decomposed into two contributions: the Kerr part and the first-order correction proportional to the scalar-field parameter $q$. The Hamilton--Jacobi equation is then solved separately for these two contributions, allowing the  action to be expressed as the corresponding perturbative sum. This procedure enables us to derive the radial and angular functions $R(r)$ and $\Theta(\theta)$ analytically and to investigate the leading-order effects of the scalar-field deformation on the photon trajectories and the resulting shadow observables.

The radial and angular functions corresponding to the Kerr metric coefficients take the following forms for null geodesics
\begin{equation}\label{R-Theta-kerr}
	\begin{split}
		&R_{Kerr}{{(r)}}=- \Delta \, [(L-a \, E)^2+C]+[E \, (r^2+a^2)-L \, a]^2,\\
		&\Theta_{Kerr}{{(\theta)}}=C+\cos^2 \theta \, [E^2 \, a^2-\frac{1}{\sin^2 \theta} \, L^2],
	\end{split}
\end{equation}
where $C$ is the Carter constant, and $\Delta$ is defined by
\begin{equation}\label{chap2-25}
	\Delta=r^2-2\, m\, r+a^2,
\end{equation}
with this definition for $R_{Kerr}{{(r)}}$ and $\Theta_{kerr}{{(\theta)}}$, then
\begin{equation}\label{ds-kerr}
	\begin{split}
		(\frac{dS_{r}}{dr})^{2}&=\frac{R_{Kerr}(r)}{\Delta^2}, \\
		(\frac{dS_{\theta}}{d\theta})^{2}&=\Theta_{Kerr}(\theta),
	\end{split}   
\end{equation}
The radial and angular functions for another part are given as follows
\begin{equation}\label{R-Theta-q}
	\begin{split}
		&R_{q}{{(r)}}= r \, (r-2\,m) \, (-C_{1}+\frac{r^3 \,  E^2}{2\,m-r}),\\
		&\Theta_{q}{{(\theta)}}=C_{1}-L^2 \, \csc^2 \theta,
	\end{split}
\end{equation}
where $C_{1}$ is a constant. 
Similar to the last part, by defining $R_{q}{{(r)}}$ and $\Theta_{q}{{(\theta)}}$, then
\begin{equation}\label{ds-q}
	\begin{split}
		(\frac{dS_{r}}{dr})^{2}&=q \, \frac{R_{q}(r)}{(r^2-2\,m\,r)^2}, \\
		(\frac{dS_{\theta}}{d\theta})^{2}&=q \ \Theta_{q}(\theta),
	\end{split}   
\end{equation}

Finally, the geodesic equations $\dot{r}$ and $\dot{\theta}$ of rotating FJNW metric can be obtained in the following forms
\begin{equation}\label{drdtheta1}
	\dot{r}=\frac{\sqrt{\frac{R_{Kerr}(r)}{\Delta^2}+q \, \frac{R_{q}(r)}{(r^2-2\,m\,r)^2} }}{g_{rr}},
\end{equation}
\begin{equation}\label{drdtheta2}
	\dot{\theta}=\frac{\sqrt{\Theta_{Kerr}(\theta)+q \ \Theta_{q}(\theta)}}{g_{\theta\theta}}, 
\end{equation}
according to relations $ \frac{\partial S}{\partial x^{\mu}}= p_{\mu}$, $ p_{\mu}= g_{\mu\nu} \, \dot{x}^{\nu}$, and defining

\begin{equation}\label{dS}
	\begin{split}
		(\frac{dS_{r}}{dr})^{2}&=\frac{R_{Kerr}(r)}{\Delta^2}+q \ \frac{R_{q}(r)}{(r^2-2\,m\,r)^2},  \\
		(\frac{dS_{\theta}}{d\theta})^{2}&=\Theta_{Kerr}(\theta)+q \ \Theta_{q}(\theta).
	\end{split}
\end{equation}  
using Eqs.(\ref{ds-kerr}) and (\ref{ds-q}). Setting $q=0$ in Eqs. (\ref{drdtheta1}) and (\ref{drdtheta2}) leads to the geodesic equations of Kerr metric.

In \cite{shadow-rotating-fjnw} a rotating metric that does not satisfy Einstein's equations was used and the function $S$ was not properly separated into fully independent radial and angular components. The rotating metric that we used, satisfies Einstein's equations and one may use it to further study different aspects of the rotating FJNW metric. 

\section{Effective Potential}\label{Section4}
In this section, we study the radial behavior of the effective potential in the rotating FJNW metric given in Eq. (\ref{jnw-metric-coefficients}). The effective potential may be written as

\begin{equation}\label{eff-potential}
	\dot{r}^2+V_{eff}=0,
\end{equation}
by substituting Eq. (\ref{drdtheta1}) into Eq. (\ref{eff-potential}), one may obtain the radial behavior of the effective potential in the equatorial plane (where $\theta=\frac{\pi}{2}$). 
\begin{figure}[h]
	\centering
	(a)\includegraphics[width=0.6\textwidth]{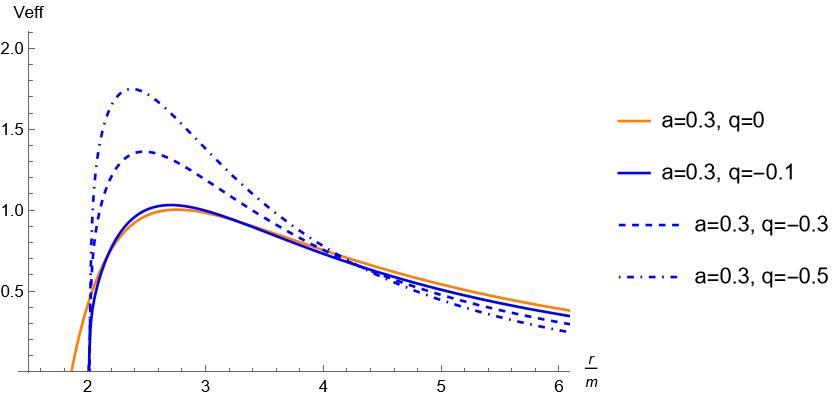}
	(b)\includegraphics[width=0.6\textwidth]{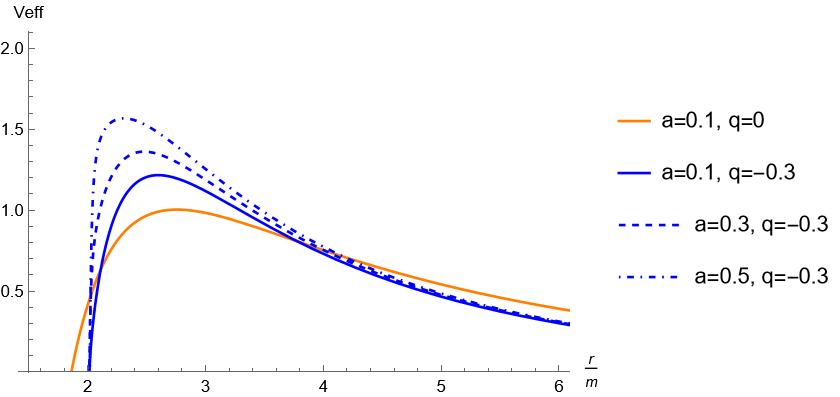}
	\caption{Radial dependence of the effective potential $V_{eff}$  for the rotating FJNW metric: (a)  for the spin parameter  $a=0.3$, and (b) for the parameter $q=-0.3$. In all plots, we set  $m=1$. Note that $q=0$ corresponds to the Kerr metric. }
	\label{fig:VeffJNWq}
\end{figure}

Figure (\ref{fig:VeffJNWq}) shows how the effective potential of photon radial motion varies with the radial coordinate for different values of $a$ and $q$. One can see the peak of the effective potential associated with the photon's path increases with increasing the absolute values of  parameters $a$ and  $ q $, indicating that photon orbits move closer to the singularity as these parameters increase. 
The innermost curve (orange) corresponds to the effective potential of a Kerr black hole ($q=0$). 

\section{Shadow}\label{Section5}

In this section, we aim to investigate the rotating FJNW metric's shadow. This analysis is at the level of null geodesic structure and their associated celestial coordinates. The shadow is therefore determined geometrically by the critical unstable photon trajectories that define its apparent boundary on the observer's celestial sky. This analysis does not include radiative transfer, emission mechanisms, or interactions of photons with the central singularity.

\begin{figure}
	\centering
	\includegraphics[width=0.9\textwidth]{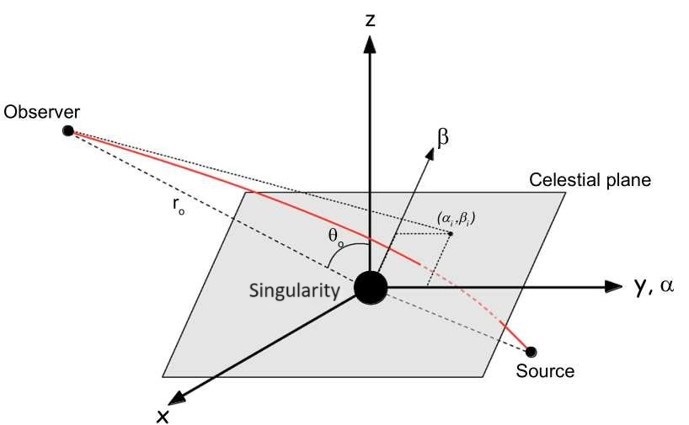}
	\caption{Schematic illustration of the distant observer’s celestial plane and celestial coordinates.}
	\label{celestial}
\end{figure}

The main objective of the present study is to investigate the morphology and deformation
of this apparent shadow boundary and its dependence on the spin parameter $a$ and the
scalar-field parameter $q$. For this purpose, one may consider three impact parameters $\xi=\frac{L}{E}$, $\eta=\frac{C}{E^{2}}$, and $\eta_{1}=\frac{C_{1}}{E^{2}}$, which characterize the boundary of the naked singularity. Now, by substituting the definitions of $\xi$ and $\eta$,  one can rewrite Eqs. (\ref{R-Theta-kerr})  as  follows
\begin{equation}\label{R-Theta-kerr1}
	\begin{split}
		&R_{Kerr}(r)= \Delta\, [-(\xi-a )^2-\eta]+[(r^2+a^2)-\xi \,a]^2,\\
		&\Theta_{Kerr} (\theta)=\eta+[a^2-\frac{\xi^2}{\sin^2 \theta}] \, \cos^2 \theta,
	\end{split}
\end{equation}
and by considering the definitions of $\xi$ and $\eta_{1}$, Eqs. (\ref{R-Theta-q}) can be rewritten in the following forms
\begin{equation}\label{R-Theta-jnw}
	\begin{split}
		&R_{q}{{(r)}}= r\,(r-2\,m)(-\eta_{1}+\frac{r^3 }{2\,m-r}),\\
		&\Theta_{q}(\theta)=\eta_{1}-\xi^2\, \csc^2 \theta.
	\end{split}	
\end{equation}
Now, to determine the unstable photon rings, the following conditions must be satisfied,
\begin{equation}\label{unstable}
	V_{eff}=0,\quad  \frac{\partial V_{eff}}{\partial r}=0, \quad  \frac{\partial^2 V_{eff}}{\partial r^2} < 0, 
\end{equation}
where $ V_{eff}=V_{Kerr}+q \, V_{q} $. Using Eqs. (\ref{drdtheta1}), (\ref{eff-potential}), and (\ref{unstable}), one can derive the impact parameters $\xi$ and $\eta$ for Kerr black hole as follows
\begin{equation}\label{eta-cesi-kerr}
	\begin{split}
		&\xi= \frac{r^2\, (-3\, m + r) + a^2\, (m + r)}{a\, (m - r)},\\
		&\eta=\frac{r^3\, (-4 \,a^2\, m + r\, (-3\, m + r)^2)}{a^2\, (m - r)^2}.
	\end{split}	
\end{equation}
Now, by using  Eqs.
(\ref{eta-cesi-kerr}) and Eq. (\ref{unstable}), we can obtain $\eta_{1}$ from the $q$-dependent part of the effective potential of rotating FJNW metric as follows
\begin{equation}\label{eta1}
	\begin{split}
		\eta_{1}=&\frac{r^3}{(2\, m - r)}-\frac{2\,a^2\, (m + r)^2\,\ln[1 -\frac{2\,m}{r}]}{(m -r)^2 }+\\
		&\frac{2\, r^2 \,(12\, m^3 - 7\, m^2 \,r - 2\, m \,r^2 + r^3) \ln[1 -\frac{2\,m}{r}]}{(m - 
			r)^2 \,(2 \,m - r)}.
	\end{split}
\end{equation}
We aim to determine the apparent shape of the shadow corresponding to the rotating FJNW, expressed through the celestial coordinates $\alpha$ and $\beta$. The coordinate  $\alpha$  represents the apparent perpendicular distance from the axis of symmetry, while $\beta$ indicates the distance relative to the projection onto the equatorial plane \cite{celestial-1, celestial-2}. Fig.  (\ref{celestial}) illustrates these coordinates.
According to the standard procedure, $r_{0}$ is the distance from the singularity to a distant observer. Celestial coordinates $\alpha$ and $\beta$ show the apparent perpendicular distance between the bright image around the naked singularity, (i) the symmetry axis and (ii) its equatorial plane projection. The angle 
$\theta_{0}$
is the angle between the singularity's rotation axis and the observer's line of sight. 
For a distant observer, the shadow’s apparent form is described by 
$\alpha$ and $\beta$ as follows
\begin{equation}\label{alpha-beta}
	\begin{split}
		&\alpha=\lim_{r_{0}\rightarrow \infty}(- r_{0}^2\, \sin \theta\, \frac{d\phi}{dr} \arrowvert_{\theta \rightarrow \theta_{0}}),\\
		&\beta=\lim_{r_{0}\rightarrow \infty}(r_{0}^2\,  \frac{d\theta}{dr} \arrowvert_{\theta \rightarrow \theta_{0}}),
	\end{split}	
\end{equation}
then, by substituting the equations of motion from Eqs. (\ref{tdot-phidot}), (\ref{drdtheta1}), and (\ref{drdtheta2}) into Eq. (\ref{alpha-beta}), it can be rewritten in the following form
\begin{equation}\label{chap2-a,b}
	\begin{split}
		&\alpha =\frac{-\xi \,\csc \theta}{ \sqrt{1 - q }},\\
		&\beta=\frac{\sqrt{\eta + a^2\, \cos^2\theta + 
				q \, \eta_{1} - \xi^2\, \cot^2\theta -  q\, \xi^2 \, \csc^2 \theta}}{ \sqrt{1 - q }}.
	\end{split}
\end{equation}
Eqs. (\ref{chap2-a,b})
reduce to the celestial coordinates of the Kerr metric when $q=0$.

\begin{figure}[h]
		\centering
		\includegraphics[width=0.6\textwidth]{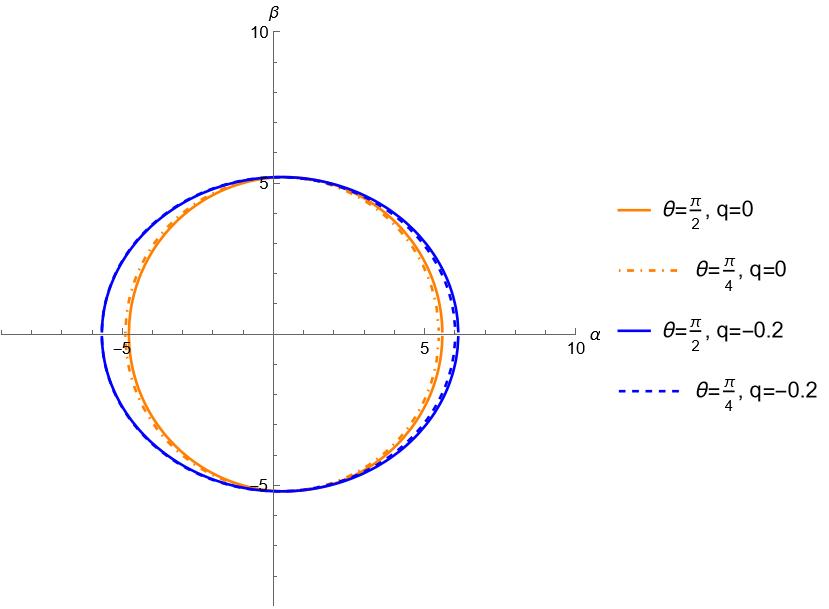}
		\caption{Shadow of the rotating FJNW metric for selected values of the angle  $\theta$, with $a=0.2$, $q=-0.2$, and $m=1$ for all diagrams. $q=0$ corresponds to the Kerr metric.}
		\label{shadowJNW-kerrpi-new}
\end{figure}

\begin{figure}[h]
		\centering
		\includegraphics[width=0.6\textwidth]{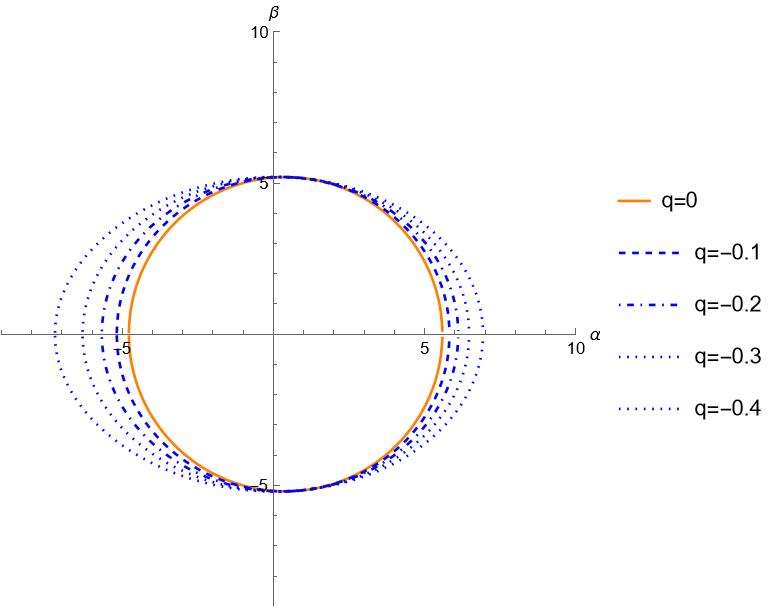}
		\caption{Shadow of the rotating FJNW metric for selected values of parameter $q$, with $a=0.2$, $\theta=\frac{\pi}{2}$, and  $m=1$ for all diagrams. }
		\label{shadowJNWq}
\end{figure}

\begin{figure}[h]
		\centering
		\includegraphics[width=0.6\textwidth]{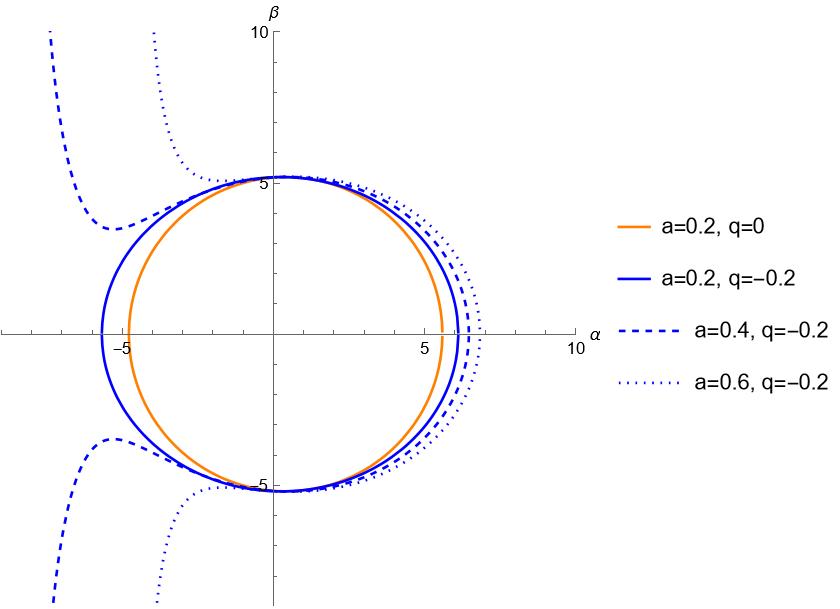}
		\caption{Shadow of the rotating FJNW metric for selected values for spin parameter  $a$, with $q=0$ and $q=-0.2$, $\theta=\frac{\pi}{2}$, and $m=1$.}
		\label{shadowJNWnew1}
\end{figure}

\begin{figure}[h]
		\centering
		\includegraphics[width=0.6\textwidth]{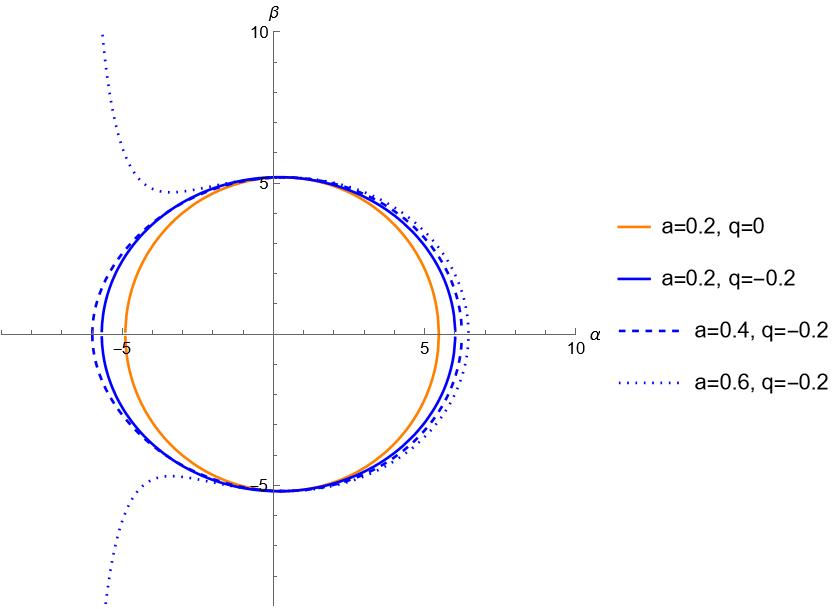}
		\caption{Shadow of the rotating FJNW metric for selected values of spin parameter  $a$, with $q=0$ and $q=-0.2$, $\theta=\frac{\pi}{4}$, and $m=1$.  }
		\label{shadowJNWnew2}
\end{figure}

Fig.  (\ref{shadowJNW-kerrpi-new}) shows the rotating FJNW metric's shadow corresponding to multiple values of the $\theta$ parameter. The innermost
curve (orange) corresponds to the Kerr black hole shadow $(q = 0)$. It is observed that the size of the shadow increases on one side as the angle $\theta$ decreases.
Fig. (\ref{shadowJNWq}) shows the rotating FJNW metric's shadow corresponding to multiple values of parameter $q$ in the equatorial plane. 
Figures (\ref{shadowJNWnew1}) and (\ref{shadowJNWnew2}) show the shadow of the rotating FJNW metric at  $\theta=\frac{\pi}{2}$ and $\theta=\frac{\pi}{4}$, respectively. As the spin parameter $a$ increases, the size of the shadow increases, and the photon ring gradually begins to open.

\newpage

\newpage

It is worth noting that the frame-dragging effects arising from the $aq$-dependent terms in the metric coefficient $g_{t\phi}$ can be investigated more carefully by expanding the metric coefficients to second order in the mass parameter $m$ and neglecting terms proportional to $a^2\,m^2$, rather than neglecting the $aq$-dependent terms. This procedure introduces additional $aq$-dependent contributions into $g_{t\phi}$, while the other metric coefficients remain unchanged.
\\
\\
Then, repeating the Hamilton–Jacobi analysis using the modified component, results in the Hamilton–Jacobi equation remains separable by neglecting $a^2 \, m^2$ terms after including these additional $aq$-dependent contributions. The additional terms in $g_{t\phi}$ modify the constant $\eta_1$, whereas the parameters $\xi$ and $\eta$ remain unchanged. Equations of motion $\dot{t}$, $\dot{r}$, and $\dot{\phi}$ also modify, while $\dot{\theta}$ remains unchanged. Consequently, the horizontal celestial coordinate $\alpha$,  remains unchanged, and as a result,  $\delta \alpha=0$. On the other hand, the vertical celestial coordinate $\beta$ is affected by an additional $q$-dependent term, $-q \, (\xi^2+1)\csc^2\theta$, which appears as an additional contribution in the numerator of the fraction under the square root defining $\beta$, together with the modification of the separation constant $\eta_1$ which leads to  $\delta\beta\neq0$.
\section{Gauss-Bonnet Theory}\label{gauss-bonnet-sec}
One of the conventional methods for obtaining the deflection angle of light is the Gauss-Bonnet method, introduced by Carl Friedrich Gauss and Pierre Bonnet \cite{def2, def3}.
	The deflection angle of a light beam in a stationary and axisymmetric spacetime is given by
	\begin{equation}\label{def-ang}
		\delta_{\Phi}  = -  \int \int_{T} {K\, dS}+ \int _ {r_{1}} ^ {r_{2}}  \kappa_{g}\, d\ell,
	\end{equation}
	where $\kappa_{g}$ denotes the geodesic curvature, $K$ is a gaussian curvature, $dS$ is the surface element, $d  \ell$ is a length element, and $T$ represents the area of a quadrilateral embedded in curved space.
	\subsection{Gaussian Curvature}
	To obtain the value of Gaussian curvature, we will use the Gibbons-Werner method \cite{Gibbons}. 
	Assuming motion in the equatorial plane,  
	where $\theta=\frac{\pi}{2}$, 
	the metric for photons can be written as follows
	\begin{equation}\label{eq-for-light}
		-A(r) \, dt^2+B(r) \, dr^2+r^2 \, C(r) \, d\phi^2=0.
	\end{equation}
	By multiplying the  above  equation by $A(r)^{-1}$, one can define
	\begin{equation}\label{g-bar}
		\bar{g}_{rr} =\frac{B(r)}{A(r)}, \quad
		\bar{g}_{\phi \phi}=  \frac {r^2 \, C(r)}{A(r)},
	\end{equation}
	the Gaussian curvature is defined by 
	\begin{equation}\label{Gaussian-Curvature}
		K=-\frac{1}{\sqrt{\bar{g}}}\, [\partial_{r}\,(\frac{1}{\sqrt{ \bar{g}_{rr}}}\, \partial_{r}\, \sqrt{ \bar{g}_{\phi \phi}})+\partial_{\phi}\,(\frac{1}{\sqrt{ \bar{g}_{\phi \phi}}}\, \partial_{\phi}\, \sqrt{ \bar{g}_{rr}})],
	\end{equation}
	where $\bar{g}= \bar{g}_{rr}\, \bar{g}_{\phi \phi}$.   
	Using Eqs. (\ref{jnw-metric-coefficients}), (\ref{eq-for-light}),   (\ref{g-bar}) and  Eq. (\ref{Gaussian-Curvature}), the Gaussian curvature of the rotating FJNW metric, after applying the transformation $u=\frac{1}{r}$, is given by
	\begin{equation}\label{gauss-curve-amount1}
		\begin{split}
			K=&	( \frac{-1+m\,u \, (3+2\,q+a^2\,u^2\,(2+q)+2\,a^2\,u^3\,m\,q)}{u\,(1-2\,m\,u)^2 \, (q\, \ln(\frac{1}{1-2\,m\,u})-1)^2} \\
			& + \frac{q\, \ln(\frac{1}{1-2\,m\,u}) \, (-2\,a^2\,u^3\,m+q\,(1-3\,m\,u) \,  \ln(\frac{1}{1-2\,m\,u}))}{u\,(1-2\,m\,u)^2 \ (q\, \ln(\frac{1}{1-2\,m\,u})-1)^2}) \\  
            & \times (\frac{u^2\,(2\,m\,u-1)^4 \, (q\, \ln(\frac{1}{1-2\,m\,u})-1)^2}{1+a^2\,u^2 \, (1+2\,m\,u)+q\, \ln(\frac{1}{1-2\,m\,u})} \\
            & \times \frac{1}{1-u\,(2\,m+u\,a^2)+q\,(1-2\,m\,u)\, \ln(\frac{1}{1-2\,m\,u})} )^{\frac{1}{2}}. \
		\end{split}
    \end{equation}

	We now consider the geodesic Lagrangian corresponding to Eq. (\ref{eq-for-light}), parameterized by the $\lambda$, as follows
	\begin{equation}
		\mathcal{L}= -\frac{1}{2} \, A(r) \, \dot{t}^2+\frac{1}{2} \, B(r) \, \dot{r}^2+\frac{1}{2}\, r^2 \, C(r) \, \dot{\phi}^2,
	\end{equation}
	in this expression, $A(r)$, $B(r)$, and $C(r)$ denote the metric functions.  Therefore, we have
	\begin{equation}\label{lagrange}
		\begin{split}
			&\frac{\partial \mathcal{L}}{\partial \dot{t}}=A(r) \, \frac{dt}{d\lambda}=E=const,\\
			&\frac{\partial \mathcal{L}}{\partial \dot{\phi}}=C(r) \, r^2 \, \frac{d\phi}{d\lambda}=L=const,
		\end{split}
	\end{equation}
	where $E$ and $L$ correspond to the photon’s energy and angular momentum, respectively. Also, using Eq. (\ref{lagrange}), we obtain
	\begin{equation}\label{lagrange-impact}
		\frac{A(r)}{C(r) \, r^2}\,(\frac{dt}{d\phi})= \frac{E}{L}=b^{-1},
	\end{equation}
	where $b$ denotes the impact parameter of the particle. Now, by solving $(\frac{dr}{d\lambda})^2$ via Eq. (\ref{eq-for-light}) and applying the transformation $r=\frac{1}{u}$, we obtain
	\begin{equation}
		(\frac{du}{d\phi})^2=\frac{C(r)}{B(r)} \, (\frac{C(r)}{A(r)\, b^2}-u^2),
	\end{equation}
	then,   expanding the obtained relation up to the order  $u^2$, we have
	\begin{equation}
		\begin{split}\label{d2u}
			(\frac{d^2 u}{d\phi^2})+u=& \,3\,m\,u^2+ \frac{3\,u\,a^2+ 2\,m\,q+4\,m^2\, q\,(1+2\,q)\,u}{b^2} \\
			&+\frac{m\,(9\,a^2\,(1+q)+8\,m^2\,q\,(1+3\,q\,(1+q)))\,u^2}{b^2}.
		\end{split}
	\end{equation}
	The homogeneous solution of  Eq. (\ref{d2u}) is given by
	\begin{equation}\label{solution-u}
		\frac{d^2u}{d\phi^2}+u=0,  \hspace*{3mm}  \Rightarrow  \hspace*{3mm}  u=\frac{\sin\phi}{b}.
	\end{equation}
	Also, Eq. (\ref{d2u}), expanded up to order $ \frac{1}{b^{2}}$, can be written as
	\begin{equation}\label{d2u-1}
		(\frac{d^2 u}{d\phi^2})+u-\frac{2\,m\,q}{b^2}- 3\,m\,u^2=0.
	\end{equation}
	Now, we consider $u$ in the following form
	\begin{equation}
		u=\frac{\sin\phi}{b}+F(\phi),
	\end{equation}
	therefore,   the solution of the differential Eq. (\ref{d2u-1}) is given by
	\begin{equation}\label{sol}
		u=\frac{\sin\phi}{b}+\frac{m\, (3 + 4 \,q +\cos 2 \phi)}{2\,b^2}.
	\end{equation}
	Finally, by substituting Eq. (\ref{sol}) into Eq. (\ref{gauss-curve-amount1}) and expanding the resulting expression up to $\frac{1}{b^2}$, one can rewrite Eq. (\ref{gauss-curve-amount1}) as  follows
	\begin{equation}\label{gauss-curve-amount}
		K=\frac{-4 \,b^2 + m^2 (15+32\,q + 16  \,q^2 + \cos 2 \phi)+ 
			8\, b \,m\, (1 + q)\, \sin\phi}{4\,b^2}.
	\end{equation}
	Finally, using Eq. (\ref{gauss-curve-amount}), we evaluate the integral of the Gaussian curvature associated with the rotating FJNW metric as follows
	\begin{equation}\label{int-gauss}
		\int_{0}^{\pi} {K\, d\phi}=	\frac{m \,(16 \,b \,(1 + q) + m\, \pi\, (15 +32\,q+ 16 \,q^2))}{4\, b^2}.
	\end{equation}
	\subsection{Geodesic Curvature}
\begin{figure}
		\centering
		\includegraphics[width=0.6\textwidth]{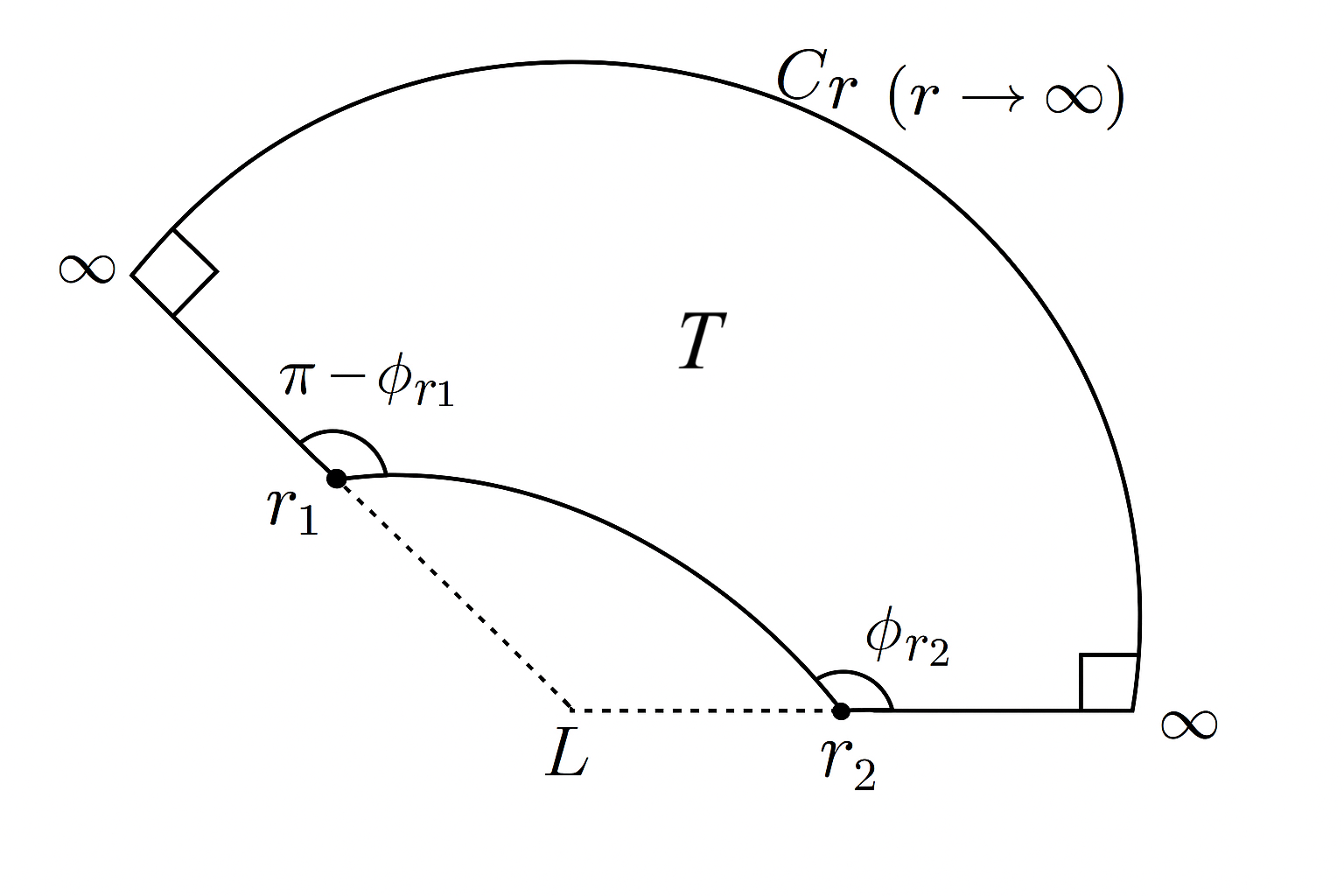}
		\caption{{Quadrilateral embedded in a curved manifold.} 
			This diagram illustrates a quadrilateral $T$ within a curved space-time which models light propagation in a non-Euclidean geometry. 
			The point $L$ represents the light source, emitting light-rays  reaching the boundary points $r_1$ and $r_2$ along a curved surface. 
			The corresponding angles $\pi - \phi_{r_1}$ and $\phi_{r_2}$ describe the orientation of the light-rays relative to the local tangent directions at those points. 
			The outer arc $C_r$ indicates the contour at infinity ($r \to \infty$), corresponding to the asymptotic boundary of the curved region.
		}
		\label{deflec2}
	\end{figure}

	The geodesic curvature represents the angular deviation of a curve from a flat surface. Also, in Eq. (\ref{def-ang}), $r_{1}$ and $r_{2}$ are two vertices of a quadrilateral in a curved space, which show the path of photons exiting from the light source to the other two vertices of this quadrilateral located at infinity (see Fig. (\ref{deflec2})).
	The geodesic curvature is obtained by the following relation
	\begin{equation}\label{geodesic-curve}  	\kappa_{g} = - (g_{rr} \, g_{\phi \phi})^{- \frac{1}{2}}  \frac{\partial D}{\partial r}, 
	\end{equation}
	where $D=-\frac{g_{t\phi}}{g_{tt}}.$ By substituting Eq. (\ref{jnw-metric-coefficients}) into Eq.  (\ref{geodesic-curve}), the geodesic curvature of the rotating naked FJNW metric can be expressed as follows
	\begin{equation}\label{kg}
		\kappa_{g}= \frac{2\,a\,m}{r^3}+O\,[m^2],
	\end{equation}
	where the term $O[m^2]$ includes higher-order terms of mass $m$, which we disregard. The result of the geodesic curvature integral, assuming a weak field regime and a low angular momentum of the rotating FJNW metric, leading to a linear consideration of photon orbits, as follows
	\begin{equation}\label{r-l}
		r=\frac{b}{\cos \theta}+O\,(m,  a), \quad
		\ell= b \, \tan \theta +O\,(m,  a),
	\end{equation}
	where $\ell $ denotes the orbital angular momentum of the photon.  Based on Fig. \ref{deflec2}, we obtain the following relations
	\begin{equation}\label{sinrs}
		\sin \phi_{r_{1}}=-\sqrt{r_{r_{1}}^2-\frac{b}{r_{r_{1}}}}, \quad
		\sin \phi_{r_{2}}=\sqrt{r_{r_{2}}^2-\frac{b}{r_{r_{2}}}}.
	\end{equation}
	Therefore, using Eqs. (\ref{kg}), (\ref{r-l}), and (\ref{sinrs}), the second term in  Eq. (\ref{def-ang}) takes the following form
	\begin{equation}\label{int-geo-curve}
		\int _ {r_{1}} ^ {r_{2}} \kappa_{g}\, d\ell=-\frac{4\,m\,a}{b^2}.
	\end{equation}
	Now, by substituting  Eqs. (\ref{int-gauss}) and (\ref{int-geo-curve}) into Eq. (\ref{def-ang}), the light deflection angle for the rotating FJNW metric can be obtained as follows
	\begin{equation}\label{def-ang-final}
		\delta_{\Phi} =\frac{4\,m}{b}+\frac{4\,m\,q}{b}+ \frac{15\, m^2\, \pi}{4\, b^2} +  \frac{8\, m^2\, \pi\, q}{b^2}+ \frac{4\, m^2\, \pi\, q^2}{b^2}-\frac{4\,m\,a}{b^2}.
	\end{equation}
	By putting $q=0$, the above solution represents the deflection angle for Kerr black hole.
	\begin{figure}
		\centering
		\includegraphics[width=0.7\textwidth]{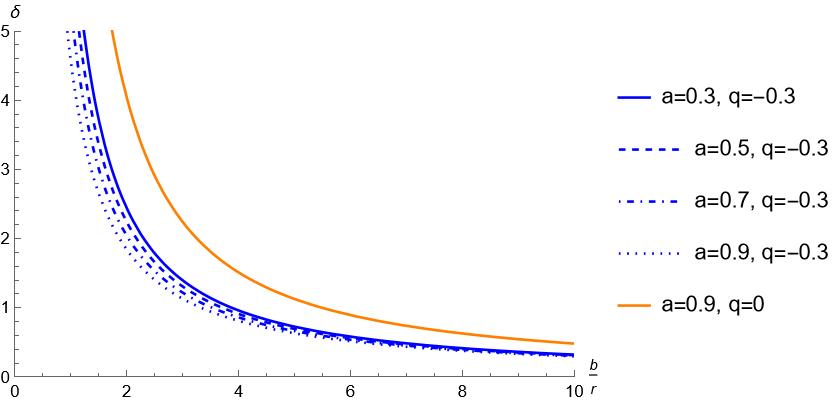}
		\caption{ Deflection angle of light of the rotating FJNW metric for constant values of parameters $b$ and $r$, and selected values of parameter $a$, with $q=-0.3$ and $m=1$. $q=0$ corresponds to the Kerr metric.  }
		\label{fig-deflection-a}
	\end{figure}
	\begin{figure}
		\centering
		\includegraphics[width=0.7\textwidth]{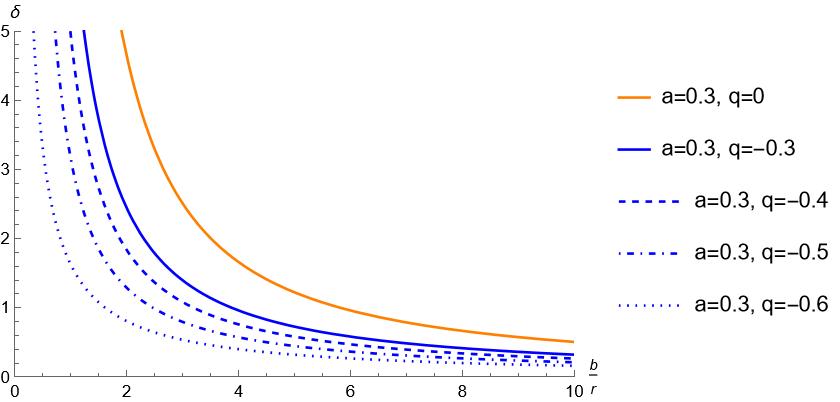}
		\caption{Deflection angle of light of the rotating FJNW metric for constant values of parameters $b$ and $r$, and selected values of parameter $q$, with $a=0.3$ and $m=1$.} \label{fig-deflection-b}
	\end{figure}  
	
	Figs. (\ref{fig-deflection-a}) and (\ref{fig-deflection-b}) show the light deflection angle in the presence of rotating FJNW metric for the constant values of impact parameter $b$ and parameter $r$. 
	In the Fig. (\ref{fig-deflection-a}), it can be observed that for a given value of the parameter $q$, by increasing the spin parameter, the light deflection angle decreases. Conversely, Fig. (\ref{fig-deflection-b}) shows that for the constant value of spin parameter $a$, the light deflection angle decreases by decreasing the parameter $q$ from $0$ to $-0.6$.
	\\
	The orange curves in Figs. (\ref{fig-deflection-a}) and (\ref{fig-deflection-b}) describe the light deflection angle in the presence of Kerr metric ($q=0$). By comparing the deflection angles of light for  Kerr (orange curve) and FJNW metrics (blue curve), it can be seen that, for the same amount of the angular momentum $ a$, the light deflection angle for rotating FJNW metrics are less than Kerr black hole.

	\section{Conclusions and Discussions}
	In this paper, we analyzed the shadow of the rotating  FJNW metric that was recently introduced in \cite{ mirza-stationary}. The trajectories of photons in the FJNW spacetime were studied via using the effective potential. It was demonstrated that as the spin parameter $a$ and the parameter $q$  increase, the circular orbits of photons around the singularity move closer to the central object (see Fig. (\ref{fig:VeffJNWq})). The equations of  motion for  photons around the rotating FJNW metric were derived analytically through the Hamilton–Jacobi differential equation and by perturbing the metric. The shadow, described in the celestial coordinate system, was determined from the geodesic equations.
	By plotting the shadow in  the celestial coordinates, we found that increasing the spin parameter $a$ and parameter $q$ leads to an increase in the size of the shadow. Also,  for   higher values of these parameters, the photon orbits begin to open up and extend their path towards infinity (Figs. (\ref{shadowJNW-kerrpi-new}) to  (\ref{shadowJNWnew2})).
	Finally,  using the Gauss-Bonnet theorem and the Gibbons-Werner method, the light deflection angle for the rotating FJNW spacetime was investigated. It was shown that  increasing the parameter $q$ leads to a larger light deflection angle, whereas increasing the parameter $a$ leads to a smaller light deflection angle in the FJNW spacetime (Figs. (\ref{fig-deflection-a}) and (\ref{fig-deflection-b})).
	\\
	
	The physical properties of the rotating FJNW spacetime as a gravitational lens can be further explored in the presence of a plasma medium and in systems containing a nearby companion or a binary configuration. Such studies would provide a more realistic understanding of lensing effects under more realistic astrophysical conditions. Additionally, obtaining more accurate solutions for the shadow of
	the rotating FJNW spacetime could extend the analysis to a wider class of exact
	stationary solutions of Einstein equations, enabling a deeper comparison between
	different gravitational models.
	
	\section{Acknowledgements}
	We acknowledge Isfahan University of Technology for the financial support that was made available to us.

\end{document}